\documentclass[12pt]{iopart}
\usepackage{mathrsfs, bm, amssymb}
\usepackage{graphicx,color}
\usepackage{iopams}
\usepackage{nameref}
\usepackage{subfig}
\usepackage{placeins}
\usepackage{float}
\usepackage{verbatim}
\usepackage{enumerate}
\usepackage{esvect}

\begin{document}

\title{Inference for interacting linear waves in ordered and random
  media}
\author{P. Tyagi$^{1,2}$, A. Pagnani$^{3,4}$,
  F. Antenucci$^{1}$, M. Ib{\'a}nez Berganza$^{5,2}$, L. Leuzzi$^{1,2}$ }

\address{$^1$ IMIP-CNR,
  Rome Unit {\em Kerberos}, Piazzale Aldo Moro 2, 00152 - Rome, Italy}

\address{$^2$ Department of Physics, {\em La Sapienza} University,
  Piazzale Aldo Moro 2, Rome, Italy}

 \address{$^3$ Department of
  Applied Science and Technology and Center for Computational
  Sciences, Politecnico di Torino, Corso Duca degli Abruzzi 24, Torino, Italy} 

\address{$^4$ Human Genetics Foundation-Torino, Via Nizza 52, Torino,
  Italy}

\address{$^5$ INFN, Gruppo Collegato di Parma, via G.P. Usberti,
7/A - 43124, Parma, Italy}

\begin{abstract}
A statistical inference method is developed and tested for pairwise
interacting systems whose degrees of freedom are continuous angular
variables, such as planar spins in magnetic systems or wave phases in
optics and acoustics.  We investigate systems with both deterministic
and quenched disordered couplings on two extreme topologies: complete
and sparse graphs.  To match further applications in optics also
complex couplings and external fields are considered and general
inference formulas are derived for real and imaginary parts of
Hermitian coupling matrices from real and imaginary parts of complex
correlation functions.  The whole procedure is, eventually, tested on
numerically generated correlation functions and local magnetizations
by means of Monte Carlo simulations.
\end{abstract} 
\maketitle

Classical $XY$ spin models with linearly interacting spins have been
subject of intensive study in statistical mechanics since the mid
$60$'s concerning the investigation of critical phenomena
\cite{Stanley68}. In particular, starting as classical lattice field
proxy for the quantum theory of the $\lambda$-transition of the Bose
condensation of optical phonons \cite{Vaks66}, and further of liquid
helium to its super fluid state \cite{Brezin82}, these models have
been employed in the theoretical description of the 2D
Kosterlitz-Thouless transition to an ordered unmagnetized spin vortex
phase \cite{Kosterlitz72,Kosterlitz73,Kosterlitz74} and, more
generally, to the study of the chirality transition
\cite{Bokil96,Kawamura10, Alba10,Obuchi13} and to the critical
behavior on random graphs, whose nature depends on their spectral
dimension \cite{Cassi92,Burioni99,Ibanez13}. Further applications can
be found to the roughening transition of the interface of a crystal in
equilibrium with its vapor \cite{Cardy96} and to synchronization
problems approached by means of the Kuramoto model \cite{Kuramoto75,Acebron05,
  Gupta14}, just to make a few examples.

Pairwise XY models can, as well, describe propagation and
amplification of linear waves in open and random media, as derived,
e. g., in Refs. \cite{Hackenbroich02,Viviescas03,Hackenbroich03} as a
classical degradation of the quantum theory for modes with overlapping
resonances.  In this framework, coupling constants yield information
about the interaction between localized (inner) modes with discrete
frequencies and radiative (outer) modes whose frequencies take values
in a continuous dominion.  In presence of relevant amount of disorder,
the refractive index strongly and inhomogeneously depends on the
spatial coordinates of the randomly placed scatterers inside the
optically active medium.  In these cases, a contribution to the
couplings comes from the spatial overlap between the electromagnetic
fields of the inner eigenmodes, modulated by a linear, inhomogeneous,
susceptibility \cite{Antenucci14,Antenucci14c}. A quantitative estimate of the
coupling coefficients, thus, yield fundamental information about the
space localization of the modes, and about the space dependence of the
optical susceptibility.  Eventually, XY pairwise models can be applied
to another problem concerning light propagation, that is, the
optimization of the transmission matrix of complex random media
\cite{Popoff10,Akbulut11}, modeled as a coupling between input and
output mode phases.

In the present work we undergo the investigation of statistical
inference techniques on XY models to provide a methodological
theoretical frame straightforwardly applicable to the above mentioned
problems with a particular focus on optics.  The developed tools can
be applied, as well, to any pairwise interacting model whose variables
can be found in $p$ states that can be considered as discrete values
of an angle, the so-called {\em $p$-clock model}
\cite{Nobre89,Ilker13}.  Recent analysis has, indeed, shown that in
the large (but not so large) $p$ limit the XY model properties are
promptly recovered for finite temperature and, further, very
interesting features arise at finite small $p$
\cite{Marruzzo14,Lupo14}. For $p=2$, eventually, one recovers the
Ising (i.e., Boolean) model for which inference studies have been carried
out in Refs. \cite{Kappen98,
  Tanaka98}.

The paper is organized as follows: in Sec.~\ref{sec:real}, we first consider the
inverse graphical problem \cite{Mezard09} for a real valued XY
model. In Sec. \ref{sec:comp}, we present the study of the generic
complex amplitude model with complex valued, and possibly disordered,
interaction couplings and we derive the relationship between
measurable correlation functions and theoretical mode couplings under
the hypothesis of a complete graph, i.e., in the so-called fully
connected mean-field limit \cite{Kappen98} were each spin is (feebly)
coupled to all the others.  In Sec.~\ref{sec:numtest} we, further,
test the obtained inference formulas on correlations numerically
generated by means of Monte Carlo simulations on different kinds of
underlying interacting networks, such as complete and sparse random
graphs.  We present our results on the efficiency of the proposed
method both in the case of ordered exchange interaction and disordered
couplings.  Eventually, in Sec.~\ref{sec:smada}, we discuss the case
where the data sets for measuring the correlation functions, from
which the couplings can be inferred, are small and in
Sec.~\ref{sec:conc} we draw our conclusions and outline the
perspectives of our work.

\section{Inference in XY model with real interaction couplings}
\label{sec:real}

 The simplest model we are going to consider consists of XY spins
 $\vec\sigma = (\cos \phi, \sin \phi) $ with angles $\phi \in
 [0,2\pi)$, pairwise real valued interaction $J_{ij}$ between sites
 $i$ and $j$ and an external field $h_i$.  Its Hamiltonian reads
\begin{eqnarray}
{\cal H}= - \sum_{(ij)} J_{ij} \cos(\phi_i - \phi_j) - \sum_i h_i
\cos(\phi_i) \label{eq:Hxy}
\end{eqnarray}
where the set of interacting pair of sites $(ij)$ is determined by
the topology of the underlying network and $J_{ij}$  is a symmetric matrix 
(the graph in undirected) whose elements can take any real
value, deterministic or randomly distributed.  The first focus of the
present work is to derive the relationship between the two-point correlation
function
\begin{equation}
C_{ij} = \langle \vec\sigma_i \cdot \vec\sigma_j\rangle =\langle
\cos\phi_i\cos \phi_j+\sin\phi_i\sin\phi_j \rangle
=\langle\cos(\phi_i-\phi_j)\rangle
\label{eq:defC}
\end{equation}
and the interaction couplings $J_{ij}$ to infer the latter for the
first.   In Eq. (\ref{eq:defC})  the average $\langle \ldots \rangle$ 
is the ensemble average over the equilibrium distribution.
Given the set of spin-spin correlation functions from
experimental data, an inverse statistical problem is setup to
investigate the interaction couplings among the spins in this
model. Such inverse problems have been widely studied for finding:
parameters of the discrete spin models using mean field theory for
complete graphs \cite{Kappen98,Tanaka98, Sessak09,Roudi09}, structural
properties of proteins from multiple sequence alignment data,
\cite{Morcos2011direct,marks2011direct, Baldassi2014PLOSONE} effective
local brain topologies from in-vivo neural recordings
\cite{schneidman2006weak}.

\subsection{Variational free energy method approach} 

Although the variational free energy method for XY model is somehow
standard \cite{Garel96}, we will briefly recall it here for fixing the
notations. We aim at finding the probability distribution
$\rho(\phi_i)$ by introducing a Lagrange multiplier $\lambda_i$ to
minimize the free energy subject to probability closure constraint
$\sum_i \int_{0}^{2\pi} d\phi_i \rho(\phi_i) = 1$. The expressions for
the internal energy $E$, entropy $S$ and free energy $F$ are the
following:
\begin{eqnarray}
E &=& -\sum_{i<j} \int_{0}^{2\pi} \int_{0}^{2\pi} d\phi_i d\phi_j \rho(\phi_i) \rho(\phi_j) J_{ij} \cos(\phi_i - \phi_j)
\nonumber
\\
&&\qquad  - \sum_i \int_{0}^{2\pi} d\phi_i \rho(\phi_i)  h_i \cos(\phi_i)
\nonumber
\\
 S &=& -\sum_i \int_{0}^{2\pi} d\phi_i \rho(\phi_i) \ln \rho(\phi_i) \quad , \qquad
\nonumber
\\
F &=& E - TS - \sum_i \lambda_i \left( \int_{0}^{2\pi}d\phi_i \rho(\phi_i)  - 1\right) 
 \end{eqnarray}
Taking the functional derivative of the free energy functional with
respect to $\rho(\phi_i)$ and setting $\delta F / \delta \rho(\phi_i)
= 0$, one finds the probability distribution $\rho(\phi_i)$.  Let us
first define:
\begin{eqnarray}
H_i^x  &=& \sum_{j} J_{ij} \langle\cos(\phi_j) \rangle +  h_i  \quad , \quad 
H_i^y = \sum_{j} J_{ij} \langle\sin(\phi_j)\rangle\quad , \quad 
\nonumber
\\
H_i &=& \sqrt{(H_i^x)^2 + (H_i^y)^2}\quad , \quad
\alpha_i = \arctan\frac{H_i^y}{H_i^x} \label{eq:4} 
 \end{eqnarray}
and the normalisation factor:
\begin{eqnarray}
 Z &=&  \int_{0}^{2\pi} d\phi_i  \exp{(H_i \cos(\phi_i - \alpha_i ))} = I_0(H_i) \label{eq:6}
 \end{eqnarray}
where $I_0(z)$ is the modified Bessel's function of the first kind.
Substituting Eqs. (\ref{eq:4}) and (\ref{eq:6}) in $\delta F / \delta
\rho(\phi_i)= 0$ one obtains
\begin{eqnarray}
 \rho(\phi_i) &=& \frac{\exp{(H_i \cos( \phi_i - \alpha_i))}}{I_0(H_i)} 
\label{eq:7}
 \end{eqnarray}

The magnetization components are, then, derived averaging $\cos(\phi_i)$ and
$\sin(\phi_i)$ over the probability measure $ \rho(\phi_i)$, yielding
\begin{eqnarray}
 m_i^x &=& \langle\cos(\phi_i)\rangle =\frac{I_1(H_i)
   \cos(\alpha_i )}{I_0(H_i)} \,\, , \,\, m_i^y =
 \langle\sin(\phi_i)\rangle = \frac{I_1(H_i)
   \sin(\alpha_i )}{I_0(H_i)} \label{eq:7a}
 \end{eqnarray}
where the modified Bessel's function of the first kind and
their derivatives  read:
\begin{eqnarray}
 I_1(z) &=&
\int_{0}^{2\pi} d\phi \cos(\phi) e^{z \cos(\phi)}
\\
\nonumber
 I^{'}_0(z) &=& I_1(z) \quad, \quad 
I^{'}_1(z) = I_0(z) - \frac{I_1(z)}{z} 
\end{eqnarray}
Correlation functions are consequently computed using the
 linear response formulas \cite{Kappen98}, i. e., 
deriving magnetizations with respect to perturbations in the external
fields $h_k$ and yielding
\begin{eqnarray}
C_{ik}^x = \frac{\delta m_i^x}{\delta h_k}
&=&
  \frac{\delta}{\delta h_k} \left \{\frac{I_1(H_i)}{I_0(H_i)} \frac{H_i^x}{H_i} \right \}
   \label{73}
\\
\nonumber
&=& \frac{H_i^x}{H_i} \left(\frac{I^{'}_1}{I_0} -
\frac{I_1^{2}}{I^{2}_0} \right)\frac{\delta H_i}{\delta h_k} +
\frac{I_1}{I_0 H_i} \frac{\delta H_i^x}{\delta h_k} -
\frac{H_i^x}{H_i^2} \frac{I_1}{I_0}\frac{\delta H_i}{\delta h_k}
\\ 
\nonumber
&=& \frac{\delta
  H_i^x}{\delta h_k} \left[(q_i - |m_i|^2) \frac{(H^x)^2_i}{H^2_i}
+ (1-q_i^2)\frac{(H^y)^2_i}{H_i^2} \right]
    \\
    &&
    \nonumber
     + \frac{\delta
  H_i^y}{\delta h_k}\frac{H_i^x H_i^y}{H^2_i}\left(2q_i - |m_i|^2-1
\right)
    \\
     \label{74} 
 C_{ik}^y
= \frac{\delta m_i^y}{\delta h_k}&=& 
 \frac{\delta H_i^y}{\delta h_k}\left[(q_i - |m_i|^2)
\frac{(H^y)^2_i}{H^2_i}+(1-q_i)\frac{(H^x)^2_i}{H_i^2}\right]
    \\
\nonumber
&&+\frac{\delta H_i^x}{\delta h_k}  \frac{H_i^x
    H_i^y}{H^2_i}\left(2q_i - |m_i|^2-1\right)
    \end{eqnarray}
where we make use of the
following substitutions:
\begin{eqnarray}
&&q_i \equiv \langle\cos^2(\phi_i)\rangle =1 - \frac{I_1}{I_0 H_i}, \quad |m_i|^2 = (m_i^x)^2 +
(m_i^y)^2
\label{eq:10}
\\
&&q_i - |m_i|^2  =\frac{I_1^{'}}{I_0} -
\frac{I_1^2}{I_0^2}    
\end{eqnarray}
We further define \begin{eqnarray}
&& \mu_i = \frac{m_i^y}{m_i^x} \\
&&  \frac{H_i^x}{H_i}
= \cos\alpha_i  = \cos[ \arctan(\mu_i )] = \sqrt{ \frac{1}{ 1 +
    \mu_i^2}} \label{eq:11} 
    \\ 
    &&\frac{H_i^y}{H_i} = \sin \alpha_i = \sin[
  \arctan(\mu_i)] = \frac{\mu_i }{ \sqrt{ 1 + \mu_i^2 }} \label{eq:11a} 
  \\
&& f^{(i)}_1(q_i,m_i) = \frac{q_i - |m_i|^2 + \mu_i^2 (1-
  q_i)}{1 + \mu_i^2} \label{eq:12}
   \\
   && f^{(i)}_2 (q_i,m_i)=
\frac{(1 + \mu_i^2)(1-q_i) +\mu_i (1-|m_i|^2)}{ 1 +
  \mu_i^2 } 
  \\
   \label{eq:12a} 
   &&g^{(i)}(q_i,m_i) = \sqrt{\mu_i}~\frac{2
  q_i - |m_i|^2 - 1 }{ 1 + \mu_i^2
} \label{eq:13}
\end{eqnarray}
Plugging the derivatives of $H_i^x$ and $H_i^y$ with respect to $h_k$
into Eqs.  (\ref{73})-(\ref{74}) and making use of Eqs. (\ref{eq:10})
- (\ref{eq:13}), we obtain a system equation for the correlation
functions $C_{ik}^{x,y}$ and the couplings matrix $J_{ik}$.  To be
compact we adopt the operator form for expressing observables
equivalent to co-ordinate form: $\bm J$ is the couplings matrix, $\bm
C^{x,y}$ the two point correlation matrices, $\mathbb{I}$ is the
$N\times N$ identity matrix and for any vector $\vec z = \{z_1,
\ldots, z_N\}$, the $N\times N$ matrix $\mathbb{I}_{\vec z}$ is
the $diag(\vec z)$ matrix, {\em i.e.} a matrix with the
elements of $\vec z$ on the diagonal and all other elements equal to
zero.  The system (\ref{73})-(\ref{74}), thus, reads
\begin{eqnarray}
\bm C^x &=&\mathbb{I}_{\vec f_1}\left[\bm J \bm C^x + \mathbb{I} \right] +
\mathbb{I}_{\vec g}  \bm J \bm C^y 
\label{eq:14}
\\
 \bm C^y &=& \mathbb{I}_{\vec g}
\left[\bm  J\bm C^x + \mathbb{I} \right] +
\mathbb{I}_{\vec f_2}
  \bm J \bm C^y \label{eq:15}
\end{eqnarray}
Let us consider the substitutions 
\begin{eqnarray}
\bm \gamma_A \equiv \bm J \bm C^x +
 \mathbb{I}  \label{eq:16} \\
\bm  \gamma_B\equiv \bm J \bm C^y   \label{eq:17} 
\end{eqnarray}
  Eqs (\ref{eq:14}) and (\ref{eq:15}) for the
correlation functions can be, accordingly, rewritten as
\begin{eqnarray}
\bm C^x &=& \mathbb{I}_{\vec f_1} \bm\gamma_A+ \mathbb{I}_{\vec g} \bm\gamma_B
 \label{eq:18} \\
\bm C^y &=& \mathbb{I}_{\vec g}  \bm \gamma_A + \mathbb{I}_{\vec f_2}  \bm\gamma_B  
\label{eq:19}
\end{eqnarray} 
To infer interaction couplings of the system we solve equations
(\ref{eq:18})-(\ref{eq:19}) in $\gamma$'s 
yielding 
\begin{eqnarray}
\bm \gamma_A &=&\mathbb{I}_{ \vec k_1} \bm C^x - \mathbb{I}_{ \vec k_2} \bm C^y  \label{eq:20} \\
\bm \gamma_B &=& \mathbb{I}_{ \vec k_3} \bm C^x - \mathbb{I}_{ \vec k_4} \bm C^y \label{eq:21} 
\end{eqnarray}
where 
\begin{eqnarray}
 \vec k_1 = \left\{\frac{f_2^{(i)}}{ \Delta_i}\right\} , \quad 
\vec k_2 = \left\{\frac{g^{(i)}}{\Delta_i} \right\}, &&\quad
 \vec k_3 = -\left\{ \frac{g^{(i)}}{\Delta_i}\right\}, 
\quad  \vec k_4 =\left\{ \frac{ f_1^{(i)}}{\Delta_i}\right\}
  \label{eq:22}
 \\
\Delta_i \equiv f_1^{(i)}f_2^{(i)} - \left(g^{(i)}\right)^2
\end{eqnarray}

 We, eventually, obtain the values of $\bm J$ by inverting
 Eq. (\ref{eq:16})
\begin{eqnarray}
\bm J   =  
  (\bm \gamma_A - \mathbb{I})(\bm C^x)^{-1}
   \label{eq:22a}
 \end{eqnarray}
and using $\bm\gamma_A$ obtained from measured correlations and
magnetizations.  Substituting for $J_{ij}$ in Eq. (\ref{eq:4}) and
using Eqs.  (\ref{eq:11})-(\ref{eq:11a}) we, moreover, obtain the
inference formula for the external field from the inferred $J$'s and
the measured $m$'s:
 \begin{eqnarray}
 \mathbb{I}_{\vec{ \mu}}\vec  h = \bm J \vec m^y -  \mathbb{I}_{\vec{ \mu}} \bm J \vec m^x
 \end{eqnarray}

\subsubsection{Zero external field}

Considering the case at $h=0$,  Eqs. (\ref{eq:4}),
(\ref{eq:7a}), (\ref{eq:10})-(\ref{eq:13}) simplify as
\begin{eqnarray}
H_i^x = \sum_{j} J_{ij} m_j^x \quad ,\quad H_i^y = 0 \quad , \quad
\alpha_i =0 \quad ,\quad H_i = H_i^x \quad  
\\
\nonumber m_i^x = \frac{I_1(H_i)}{I_0(H_i)}, \quad, m_i^y = \mu_i=0 
\\
\nonumber
f_1^{(i)}=q_i-m_i^2, \quad f_2^{(i)}=1-q_i, \quad g_i=0
\end{eqnarray}
and the  correlation function reduces, then, to
\begin{eqnarray}
C_{ik}^x &=& \frac{\delta m_i^x}{\delta h_k} \Biggr|_{\vec h=\vec 0}
  = 
  [\langle\cos^2(\phi_i)\rangle - \langle\cos(\phi_i)\rangle^2] \left(
\sum_{j} J_{ij}C_{jk}^x + \delta_{ik}\right)\label{eq:23}
  \end{eqnarray}
  Further, inverting Eq. (\ref{eq:23}), eventually
yields:
\begin{eqnarray}
\bm J &=& \mathbb{I}_{ \vv{k_1}} - \bm (C^x)^{-1} \label{eq:24}
, \qquad {\vv k_1}=\left\{\left(q_i-m_i^2\right)^{-1} \right\} 
\end{eqnarray}
In the above Eq. (\ref{eq:24}) of inferred $J$, it is worth noticing
the similarity with the expression of inferred $J$ in terms of $C$ of
the Ising model with discrete spins, see, e.g., \cite{Kappen98, Tanaka98,Roudi09}, where
 $\langle\cos^2(\phi_i)\rangle = 1$.

\section{Model with complex spins and couplings}
\label{sec:comp}
In this section, the interaction coupling matrix $J_{ij}$ is
considered to be a complex matrix consisting of a real and an
imaginary part as $J_{ij}=J^R_{ij} + \mathrm{i} J_{ij}^{I}$ and the
external field is a complex vector $h_i = h_i^{R}+ \mathrm{i}
h_i^{I}$.  This model can be a proxy for the propagation and
interaction of waves in optically active media, ordered or random,
where electromagnetic modes can be represented, in a properly defined
base of eigenvectors, by complex numbers $a_i= A_i ~e^{ \mathrm{i}
  \phi_i}$: each mode amplitude has its own magnitude $A_i=|a_i|$ and
phase angle $\phi_i=\arg(a_i)$.  Indeed, the electromagnetic field can
be decomposed, in the slow amplitude approximation \cite{Sargent78},
in terms of the complex amplitudes of the modes localized inside the
medium
\begin{equation}
\bm E(\bm r, t)=\sum_{k=1}^{N}a_k(t)~ \bm U_k(\bm r) ~e^{\mathrm{i} ~\omega_k t + \phi_k} + c.c.
\end{equation}
where the frequencies $\omega_k$ take values on a discrete dominion
and $\bm U_k$'s are the eigenvectors of the eigenmodes in some given
basis allowing for a decomposition between inner and outer modes by
means of Feshbach projectors~\cite{Hackenbroich03}. We stress that, in
cavities with non-negligible leakages or cavity-less light scattering
random media, the $N$ modes indicated in the sum are by no means
a complete basis, but they are the subset made of purely localized
modes amplified inside the cavity.
  
Such modes can be proven to display a stochastic dynamics governed by
a quantum Langevin dynamics \cite{Viviescas03}. In the classical limit
such evolution is proved equivalent to the master equation for the
density of states \cite{Hackenbroich03} and in terms of complex
amplitudes takes the form
  
\begin{eqnarray}
\dot a_n(t)&=& - \sum_m {\cal J}_{nm} a_m +\eta_n(t)
\label{eq:Lang_a}
=\frac{\partial {\cal H}}{\partial a^*_m} + \eta_n(t)
\\
\nonumber
&& \langle \eta_n(t)\rangle=0, \qquad \langle \eta_n^*(t)\eta_{n'}(t')\rangle\simeq 2 n_{\rm th}\delta_{nm} \delta(t-t')
\end{eqnarray}
where, for large enough heat-bath temperature, the thermal number of
photons in the classical regime is $n_{\rm th}\propto T$ and the
complex valued white noise is approximated as uncorrelated on
different states \cite{Gordon02,Angelani06}. In general, though, we
recall that non-diagonal covariances in the space mode can be non-zero
for Markovian dynamics \cite{Hackenbroich03}. The linear non-diagonal
coupling ${\cal J}_{nm}$ between modes is the dumping matrix,
associated to the openness of the optical cavity, due, e. g., to
leakages in standard lasers \cite{Fox68} or to the cavity-less
structure of the scattering region of the optically active material in
random lasers \cite{Wiersma08, Ghofraniha15, Eremeev11,Antenucci14}.  
The total power taken by
the system is a constant ${\cal E}$ that, rescaling the amplitudes as
$ a_n \to a_n/\sqrt{\omega_n}$, can be expressed as a simple spherical
constraint
\begin{equation}
{\cal E}=\epsilon N=\sum_{n=1}^N |a_n|^2
\label{eq:spher}
\end{equation}

The static properties of the above dynamics can be derived by studying
the Hamiltonian 
\begin{eqnarray}
{\cal H}[\{a\}] &=& - \sum_{(ij)} a_i J_{ij} a_j^{*} - \sum_i h_i a_i^{*}
\end{eqnarray}
where $J_{ij}$ includes the dumping ${\cal J}_{ij}$, incorporating the
inner-outer modes interaction, and, possibly, also includes the
spatial overlap of the eigenmodes modulated by an inhomogeneous
dielectric constant. We add a complex external field $h_i$ for
generality. In the present work, we assume $J_{ij}$ to be Hermitian,
thus $J^R_{ij}=J_{ji}^R$, and $J_{ij}^I=-J_{ji}^I$.

The dynamics of mode phases changes at a much faster time scale than
mode amplitudes \cite{Conti11}.  Moreover, in presence of a large
number of modes and a not too diluted interaction network, in a wide
variety of systems intensity equipartition occurs among all modes,
i.e., $|a_i|\simeq 1$ \cite{Antenucci14b}, trivially satisfying
constraint (\ref{eq:spher}). In particular, pairwise interacting mode systems,
 in any graph topology, always display intensity equipartition in all thermodynamic phases. 
One can thus work in the so-called {\em
  quenched amplitude} approximation, that is, amplitudes are further
taken as quenched and incorporated in the $J$'s, yielding the
Hamiltonian:

\begin{eqnarray}
{\cal  H}[\{\phi\}] &=& - \sum_{(ij)} \left[J^R_{ij} \cos(\phi_i - \phi_j)  +  J_{ij}^{I} \sin (\phi_i - \phi_j)\right] 
\nonumber \\
&&
- \sum_i 	\left[ h^R_i \cos (\phi_i) + h_i^{I} \sin (\phi_i)\right]
 \label{eq:25}
 \end{eqnarray}
 
Though derived in terms of light modes interacting in an optical open
cavity, we stress that the above Hamiltonian generically describes any
linear wave system. Indeed, it can, e. g., describe  a class of
 optimization problems where the variable $\phi$ represents the phase of 
 a pixel of the incoming/outcoming light
propagating through a random medium, including disordered optical
fibers, and $J_{ij}$ represents the transmission matrix
\cite{Popoff10,Akbulut11}.

The same variational procedure of Sec. \ref{sec:real} is applied to
the complex system modeled by Eq. (\ref{eq:25}).  The following
substitutions for $H_i^x$ and $H_i^y$ are considered, in place of
Eq. (\ref{eq:4}), in order to calculate the probability distribution
$\rho(\phi_i)$:
 \begin{eqnarray}
H_i^x &=& \sum_{j} J^R_{ij} \langle  \cos(\phi_j) \rangle - \sum_{j} J_{ij}^{I} \langle
  \sin(\phi_j) \rangle +  h^R_i \label{eq:26} \\
H_i^y &=&  \sum_{j} J^R_{ij} \langle  \sin(\phi_j) \rangle + \sum_{j} J_{ij}^{I} \langle  \cos(\phi_j) \rangle + h_i^{I} \label{eq:27}
\end{eqnarray}
It is found that the structure of $\rho(\phi_i) $ remains the same as
in the case of real-valued $J$'s, cf. Eq. (\ref{eq:7}), though the
components $ H_i^x ,H_i^y$ and $H_i$ adorn different expressions.

\subsection{Correlation functions and inference formulas}
In this section, equations for the correlation functions are
derived. In the presence of complex fields, four correlation functions
are found from the differentiation of magnetizations with respect to
both components of the fields externally acting on the system. Using
the expressions of $H_i^x / H_i$ and $H_i^y/H_i$ as in
eqs. (\ref{eq:11})-(\ref{eq:11a}), magnetizations in Eq.
(\ref{eq:7a}) can be written in the following form,
\begin{eqnarray}
m_i^x &=& \frac{I_1(H_i)}{I_0(H_i)} \frac{H_i^x}{H_i}\quad,  \qquad \label{53}
m_i^y = \frac{I_1(H_i)}{I_0(H_i)}\frac{H_i^y}{H_i} 
\end{eqnarray}
To derive correlation functions we use the following linear response
relations,
\begin{eqnarray}
C_{ik}^x = \frac{\delta m_i^x}{\delta h_k^{R}}\quad,  \qquad
\tilde{C}_{ik}^x = \frac{\delta m_i^x}{\delta h_k^{I}}\quad, \qquad
C_{ik}^y = \frac{\delta m_i^y}{\delta h^R_k}, \qquad
\tilde{C}_{ik}^y = \frac{\delta m_i^y}{\delta h_k^{I}} \label{eq:28a} 
\end{eqnarray}

Performing the above derivatives as in Sec. \ref{sec:real}, we find
equations for the correlation functions in matrix form as following
\begin{eqnarray}
\bm C^x &=& \mathbb{I} _{\vec f_1} \left[ \bm J^R \bm C^x - 
  \bm J^{I} \bm C^y + \mathbb{I} \right] 
   +  \mathbb{I}_{\vec g} \left[ \bm J^R
  \bm C^y +  \bm J^I \bm C^x\right]
   \label{eq:29}
  \\
   \bm{\tilde{C}}^x &=& \mathbb{I}_{\vec f_1} \left[\bm J^R
  \bm{\tilde{C}}^x - \bm  J^{I} \bm{\tilde{C}}^y \right] +
\mathbb{I}_{\vec g} \left[\bm J^R \bm{\tilde{C}}^y + \bm J^{I}
 \bm{ \tilde{C}}^x + \mathbb{I} \right]
  \label{eq:30} 
 \\ 
  \bm C^y &=& \mathbb{I}_{\vec g}  \left[\bm J^R \bm C^x - \bm J^I
 \bm C^y + \mathbb{I} \right] +\mathbb{I}_{\vec f_2} \left[\bm J^R \bm C^y
  + \bm J^{I} \bm C^x \right] 
  \label{eq:31} 
  \\ 
{\bm {\tilde{C}}}^y
&=& \mathbb{I}_{\vec g} \left[\bm J^R {\bm{\tilde{C}}}^x - 
 \bm J^{I} {\bm{\tilde{C}}}^y \right] +
 \mathbb{I}_{\vec f_2} \left[ \bm J^R
  {\bm {\tilde{C}}}^y + \bm J^{I} {\bm{\tilde{C}}}^x +
\mathbb{I} \right] 
\label{eq:32}
\end{eqnarray}
These are decoupled two-by-two and the two subsystems are not
independent, but equivalent to each other. Therefore, to obtain the inference formulas for $\bm J$
we can simply solve Eqs. (\ref{eq:29}) and (\ref{eq:31}).  To invert them let
us first define:
\begin{eqnarray}
\bm\Gamma_A &\equiv& \bm J^R \bm C^x - 
 \bm J^{I} \bm C^y + \mathbb{I} 
 \label{eq:GA} 
 \\
\bm\Gamma_B&\equiv &\bm J^R
\bm  C^y + \bm J^{I}\bm C^x 
\label{eq:GB}
  \end{eqnarray}
Substituting  into  eqs. (\ref{eq:29}), (\ref{eq:31}) and solving for $\Gamma$'s
we obtain 
\begin{eqnarray}
\bm \Gamma_A &=&\mathbb{I}_{ \vec k_1} \bm C^x - \mathbb{I}_{ \vec k_2} \bm C^y 
 \label{eq:37} \\
\bm\Gamma_B &=& \mathbb{I}_{ \vec k_3} \bm C^x - \mathbb{I}_{ \vec k_4} \bm C^y 
\label{eq:38} 
 \end{eqnarray}
where coefficients $\mathbb{I}_{ \vec k_a}$ are given in Eq.
(\ref{eq:22}). Now, after obtaining $\Gamma_A$ and $\Gamma_B$ in terms
of measurable quantities, we get back to eqs. (\ref{eq:GA}) and (\ref{eq:GB})
and solve them to extract interaction couplings, yielding the main
equations of our work:

\begin{eqnarray}
\bm J^R =  \left[ \left(\bm\Gamma_A- \mathbb{I} \right)
\left(\bm C^y\right)^{-1} + \bm \Gamma_B \left(\bm C^x\right)^{-1}
\right]
\left[ \bm C^x \left(\bm C^y\right)^{-1} + \bm  C^y \left(\bm C^x\right)^{-1} )
\right ] ^{-1}
 \label{eq:JRfinal}\\
\bm J^I = \left[ -\left(\bm \Gamma_A+ \mathbb{I} \right) \left(\bm
  C^x\right)^{-1} + \bm \Gamma_B \left(\bm C^y\right)^{-1}\right]
\left[ \bm C^x \left(\bm C^y\right)^{-1} + \bm C^y \left( \bm
  C^x\right)^{-1} \right]^{-1} \label{eq:34}
\end{eqnarray}

Eventually, to infer the external field values from the inferred $\bm J$ values and the measured magnetizations, using $H_i^x = {H_i^y}/{\mu_i}$ in Eqs. 
(\ref{eq:26})-(\ref{eq:27}), we obtain  
\begin{eqnarray}
\mathbb{I}_{\vec \mu} ~\vec{h}^R -\vec{ h}^I &=& (\vec m^x +
\mathbb{I}_{\vec \mu}~ \vec m^y) \bm J^I \label{eq:41}
\end{eqnarray}

\section{Numerical tests }
\label{sec:numtest}
To verify the efficiency of the predictions of the inference method
derived above, we present our tests on data provided by means of Monte
Carlo simulations of models exactly given by Eqs. (\ref{eq:Hxy}) and
(\ref{eq:25}) with known couplings and fields and defined on different
topologies.

We will first consider data generated from finite size models on fully
connected long-range models, where each spin is connected to all the
others and each coupling $J_{ij}$ is vanishingly small to guarantee
thermodynamic convergence.  These are the hypothesis at the ground of
the variational approach followed in deriving
Eqs. (\ref{eq:JRfinal}-\ref{eq:34}).  We will consider both the case of
ordered $J_{ij}=J$ and randomly distributed $J_{ij}$ with Gaussian
probability of mean zero and variance one.

We will afterwards consider data generated on models defined on 
Erdos-Renyi (ER) sparse graphs \cite{Newman03} where the connectivity is
randomly distributed according to the Poisson Distribution
\begin{equation}
P(k) = \frac{e^{-c}
  c^k}{k!}
  \label{eq:Poi}
\end{equation}
 Once again,
both deterministic and quenched disordered interaction couplings are
considered.

The values of the inferred matrix are, eventually, compared to those of the
original matrix.  
 All the data, namely,
correlation functions $C^X, C^Y$, magnetizations $m_x, m_y$ and $q$
and $|m|^2$ used for this analysis are computed from thermalized data
produced using Monte Carlo simulations with the parallel tempering
algorithm. 
For each case, 
we display (i) the comparison of the whole range of sorted original and inferred couplings,
 (ii) the comparison of inferred to original couplings to one single site 
and (iii) the sensitivity plots for true positive (fraction of original non-zero couplings inferred to be non-zero) against
the number of predicted connections, yielding  an insight into the
topology of the graph.

\subsection{Ordered couplings on complete graph}
\begin{figure}[b!]
 \includegraphics[width=0.99\linewidth]{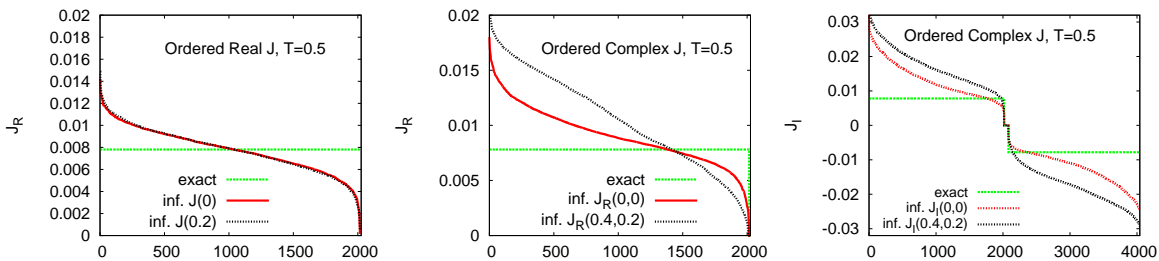}
 \caption{Sensitivity plot of the sorted inferred interaction
   couplings in the complete graph at $T=0.5$ for an ordered system of $N=64$ variables.  
   Left:
   Ordered, only real $J$ for $h=0, 0.2$. Right: $J_R$ and $J_I$
   for $\bm h = (h_R,h_I)=(0,0) $ and for $\bm h=(0.4,0.2)$.  }
          \label{fig:fc_ord_J}
\end{figure}

For the fully connected case the entire analysis is shown at $T=0.5$,
in the 
ferromagnetic phase.
In Fig. \ref{fig:fc_ord_J} we display the comparison between the
original (dotted/green lines) and the inferred (continuous/red and black lines)
couplings sorted by magnitude in a system of $N=64$ modes and
$N(N-1)/2=2016$ independent couplings.  We consider three cases. On
the left panel purely real, ferromagnetic $J_{ij}=J_{ji}=J$ couplings
are plotted in a zero external magnetic field, $h=0$, and for $h=0.2$.
To infer the values of $J_{ij}$ we used Eq. (\ref{eq:JRfinal}) that
fairly predicts the initial couplings even though the equations used
are of the general case, i.e., complex Hermitian $J$'s that can take
any value.  Using, instead, Eq. (\ref{eq:22a}) or Eq. (\ref{eq:24}),
focused on the specific cases of purely real couplings and zero-field,
no difference is appreciated down to the third digit.

In the mid and right panels, we plot real and imaginary inferred $J$'s with
zero fields and for $h_R=0.4$, $h_I=0.2$.  The quality of the
prediction is comparable with the purely real case.

    \subsection{Disordered couplings on complete graph}
    
\begin{figure}[t!]
  \includegraphics[width=0.99\linewidth]{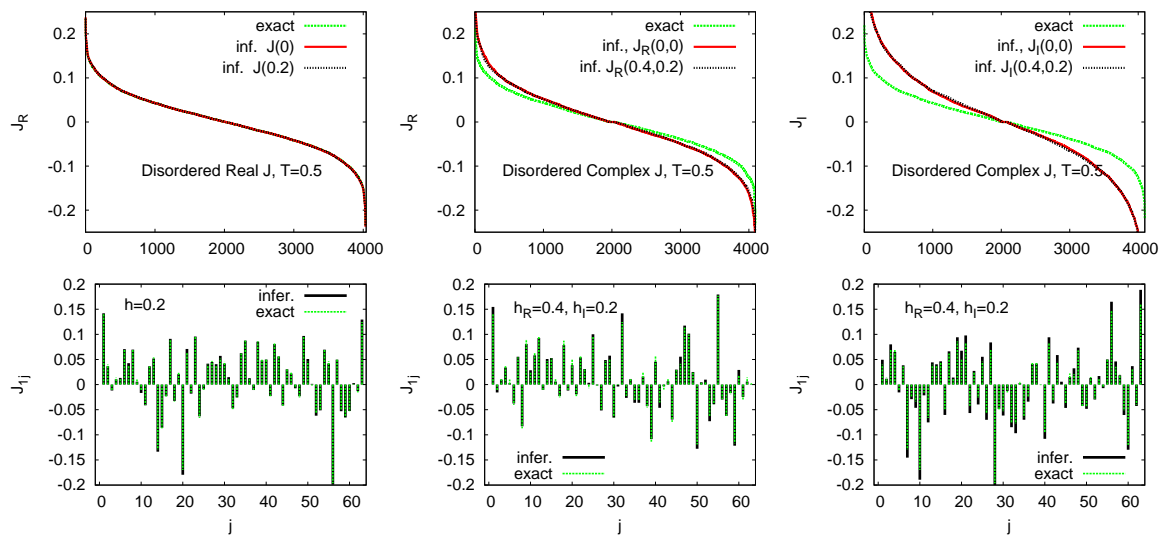}
  \caption{Sensitivity plot of the sorted inferred interaction
   couplings in the complete graph  with $N=64$ and quenched disorder at $T=0.5$, in the spin-glass phase. 
   Left: Purely real valued $J$ for $h=0,
    0.2$. Right: Real and imaginary parts of the 
    complex disordered couplings, $J_R$ and $J_I$ for $\bm h =
    (h_R,h_I)=(0.4,0.2)$. Bottom panels: original and inferred couplings to a single site.}
    \label{fig:fc_dis_J}
 \end{figure}

  We, then, inferred couplings from correlation functions and
  magnetizations generated in a system of $N=64$ spins where $J_R$ and
  $J_I$'s are originally generated by means of a Gaussian random
  distribution of mean zero and variance one.
In the left panels (top and bottom) of Fig. \ref{fig:fc_dis_J} the case of purely real couplings
is exposed, both in $h=0$ and $h=0.2$. No difference is appreciated
between these two cases and the magnitudes of both are about 
the magnitude of the  original couplings. As detailed in the bottom panel for couplings
to a specific site, inferred $J$'s faithfully predict sign and
 magnitude of the original ones.
 The center and right panels display the behavior of real and
 imaginary part of a system with complex couplings both in absence and
 presence of external fields.  Again, the presence of external fields
 do not alter the inference predictions and signs and 
 magnitude of original couplings are correctly predicted.
 
 \subsection{Ordered couplings on sparse random graph }
Next, we show the analysis for the case where the $J$ matrix is
diluted, though the formalism developed in this work is rigorous for
fully connected systems and not for sparsely connected ER graphs.  The
connectivity probability is randomly distributed according to
Eq. (\ref{eq:Poi}) with average connectivity $c=6$. Data shown are for
$N=256$ spins at temperature $T = 2.5 > T_c \sim 1.95$ 
 for the
  systems with complex Hermitian couplings, cf. Eq. (\ref{eq:25}) and
  $T=0.75$ (here $T_c=2.958$) for systems with purely real couplings,
  given by Eq. (\ref{eq:Hxy}).

  \begin{figure}[t!]
  \includegraphics[width=0.99\linewidth]{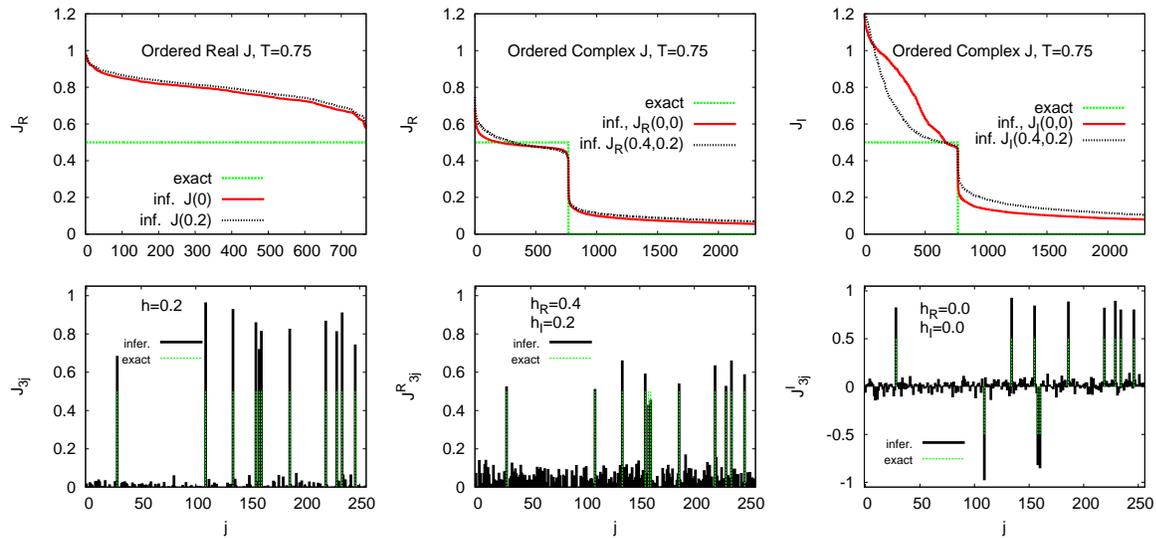}
 \caption{Sensitivity plot for the inferred couplings of an original random ER graph of $N=256$ nodes with 
 fixed deterministic values of $J$ at $T=0.75$. Left: Real valued $J$'s for the largest entries (first 
 $N~ c/2$ elements, $c=6$). Right: Real and imaginary part of complex couplings (first $3/2 N~c$ entries).
 Bottom panels: original and inferred couplings to a single site.}
\label{fig:sr_or}
 \end{figure}

\begin{figure}[t!]
\includegraphics[width=0.99\linewidth]{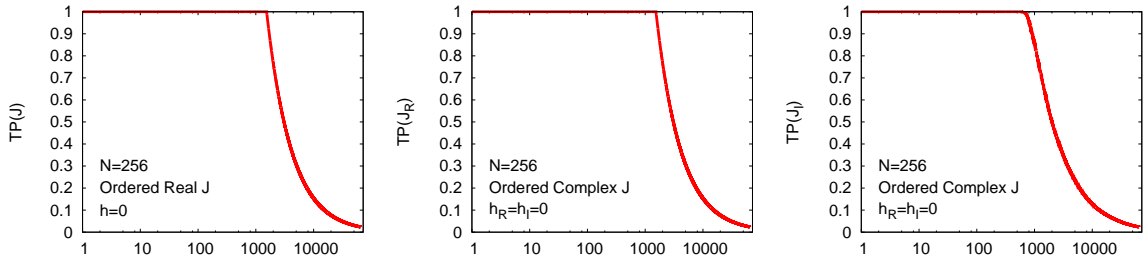}
   \caption{True positive plots for the  ER sparse random graph with average connectivity $c=6$,  size $N=256$ and with deterministic couplings $J=1$.}
\label{fig:tp_sr_or}
\end{figure}

In Fig. \ref{fig:sr_or} we display the comparison of the inferred and
the original $J$'s by means of Eq. (\ref{eq:JRfinal}). In the left panels
original couplings are all real and, when non-zero, all equal to each
other. Each site is connected to a finite, $N$-independent, number of
others, $c=6$ in the average.  The analysis gives correct indication
for non-zero $J_R$'s both in absence and presence of external fields. 
However, our method always provide non-zero (though small)
predictions for all couplings. Indeed, 
the  true
positive plots, cf. Fig. \ref{fig:tp_sr_or},
decay down to zero only gradually, not sharply, quantifying the wrong
predictions.Even though all non-zero elements of the matrix have been
predicted correctly, for every zero element the formalism does not
predict exact zero, bringing down the score
of true positive.

The same situation arises for complex couplings, where
 rather good estimates of non-zero $J^R$ and $J^I$ entries is provided, including a sharp 
 decrease of the value of the inferred couplings at the sorted coupling $N~c/2$, cf. 
 top panels in Fig. \ref{fig:sr_or}. This is, though, contrasted 
 by the rather poor estimate of zero couplings. As confirmed by the true positive
plots in Fig. \ref{fig:tp_sr_or}, 
the right panels of Fig. \ref{fig:sr_or}  show that  zero 
couplings, i. e., those beyond the $N~c/2$-th coupling, 
are inferred to acquire a non-zero value. 
Zero original couplings are not reproduced at all in the
sparse case.

\subsection{Disordered couplings on sparse random graph}  
\begin{figure}[t!]
  \includegraphics[width=0.99\linewidth]{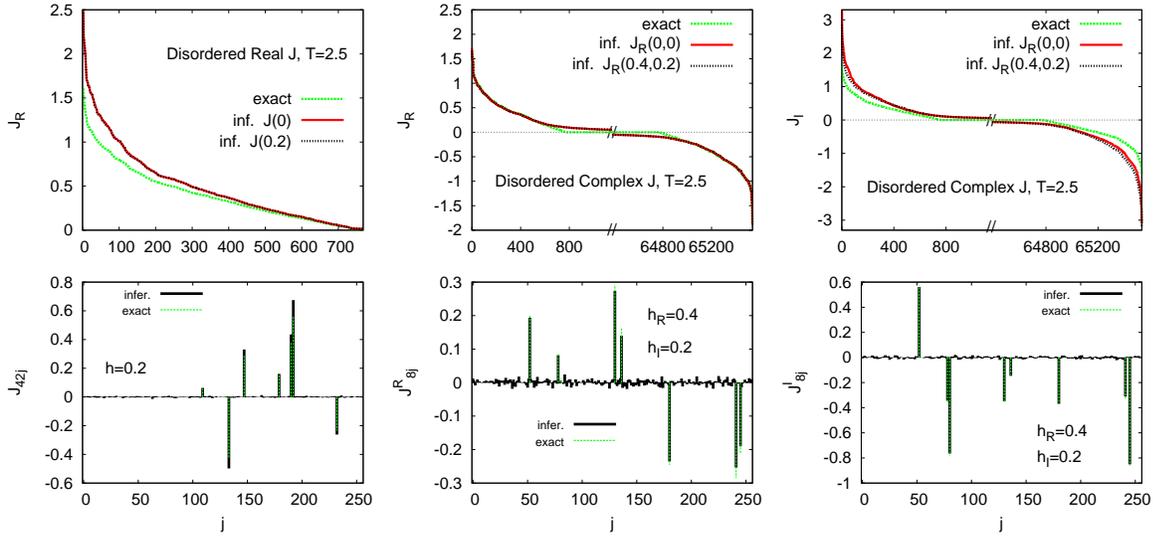}
  \caption{Sensitivity plot for the inferred couplings of an original random ER graph with 
Gaussian distributed random couplings (zero mean, unitary variance) at $T=2.5$ and with $N=256$. 
Left: Real valued $J$'s for the largest entries (first 
 $N~ c/2$ elements). Right: Real and imaginary part of complex couplings (first and last $3/4 N~c$ entries). Bottom panels: original and inferred couplings to a single site.}
\label{fig:sr_dis}
   \end{figure}

\begin{figure}[t!]
 \includegraphics[width=0.99\linewidth]{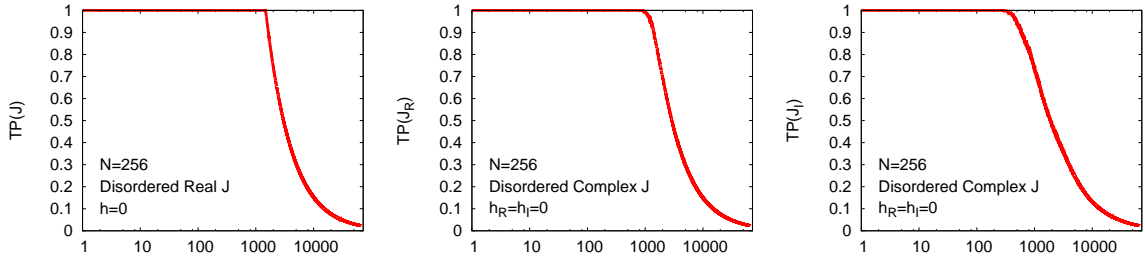}
  \caption{True positive plots for the  ER sparse random graph with average connectivity $c=6$,  size $N=256$ and with Gaussian random distributed couplings (zero mean, unit variance). Left:   True positive of a system with purely real couplings and zero external field. Mid and Right: True positive of the real and imaginary parts of the couplings in a system with zero field.}
  \label{fig:tp_sr_dis}
  \end{figure}

   The inference maintains the same quality also in the case of random values of the couplings.
   In Fig. \ref{fig:sr_dis}, left panels, we display the case of a ER random graph whose couplings have Gaussian distributed real values, with average zero and variance equal to one. The top figure in the sensitivity plot for the first $N~c/2$ couplings, with and without external field, compared to the original disordered coupling values.
   The bottom panel show the comparison between original and inferred couplings to the graph node $3$, to exemplify that: (i) all original non-zero couplings are well reproduced and discriminated in the inference procedure
   and (ii)  all inferred couplings are non zero, also those corresponding to missing original couplings, though the latter acquire a rather small value in comparison to the inferred true links.  
   The same analysis is illustrated in the mid and right panels of Fig. \ref{fig:sr_dis} for the real and imaginary part of a system   with Hermitian couplings. In the sensitivity plot the first and last $3~N~c/2$ couplings are reported and compared to the original ones,
   signaling that the inference quality is very good, though non-zero couplings are inferred to have a small non-zero value. In the bottom panels couplings to one node are displayed. 
   This is confirmed 
   in Fig.
  \ref{fig:tp_sr_dis} where the true positive curve is shown to decrease sharply after the last non-zero coupling
  but still is non-zero for all $J_{ij}$ matrix entries in all considered cases.
   
\newpage
  \section{Small data size behavior}
 \label{sec:smada}
\begin{figure}[t!]
 \includegraphics[width=0.99\linewidth]{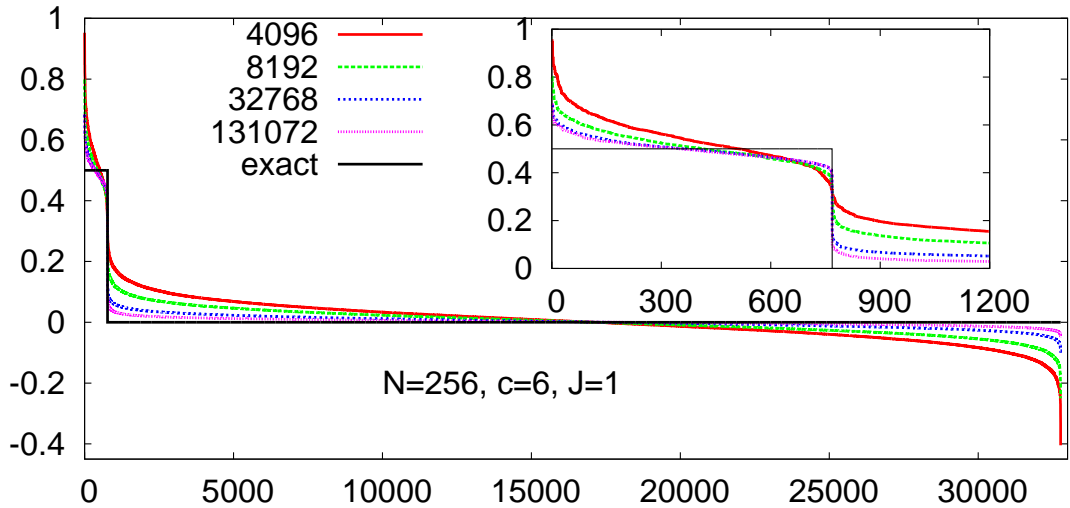}
 \caption{Sensitivity plot of data generated from an ER graph of
   $N=256$ sites and mean connectivity $c=6$ with real ordered couplings and
   no field. The behavior of the original network is marked as a full
   black step curve. Correlation functions averaged over,
   respectively, $4096, 8192, 32768$ and $131072$ data from Monte Carlo simulations are
   considered. In the inset a detail of the absolute values of the
   bonds around the last non-zero original coupling ($N~c/2=768$) is
   provided. }
\label{fig:FS}
   \end{figure}

 In this part, we show how the quality of inference is deteriorated as
 the number of measurements composing the data set used to calculate
 correlations decreases. In the main figure \ref{fig:FS}, the entire
 sorted $J$ matrix is shown and in the inset the absolute value of the
 first $2000$ elements are shown. We see that the sensitivity plot
 remains the qualitatively the same for all data sets, but the
 transition from non-zero to zero couplings becomes sharper and
 sharper as the data size increases, yielding evidence for an
 underlying sparse graph.

\section{Conclusions}
\label{sec:conc}

In the present paper we have derived an inference procedure to
determine the coupling constants of pairwise interacting systems with
continuous $XY$ spins, complex interactions and complex external
fields.

For testing the analytic inference approach we have applied it to data
numerically generated by means of Monte Carlo simulations at
equilibrium and we have compared the inferred values of the coupling
constants to the ones of the simulated system. We considered models
with disorder in the coupling values and models with disorder in the
coupling connectivity, studying both complete and sparse random graphs
with both deterministic and quenched disordered couplings.  The
inferred couplings turn out to reproduce original ones in an excellent
way in fully connected models, that is, under the conditions at the
ground of the theoretical derivation of inference formulas
Eqs. (\ref{eq:22a}, \ref{eq:JRfinal}, \ref{eq:34}).  Also when applied
to sparse random graphs, though, the quality of the inference is of a
high standard.  The only problem arises in the wrong representation
for missing links that acquire always non-zero value in the inference
procedure. Their values, actually, decrease with increasing data size, but do not reach zero even 
for very large data sizes, cf. Fig.  \ref{fig:FS}.
Else said, the true positive curve is always non-zero for
all couplings.  The value of false positive inferred couplings turns
out, though, to be systematically much smaller than the value of true
positive bonds, with a sharp quantitative distinction between the two.

In the field of random photonics, the reported method to
quantitatively infer coupling constants from experimental data would
allow to obtain estimates of the effective damping interaction between
localized modes in a random medium mediated by radiative modes
\cite{Hackenbroich02,Viviescas03,Hackenbroich03} in an open cavity, to
extrapolate the magnitude of the optical-response-modulated spatial
overlap between those modes \cite{Antenucci14} and, eventually to
obtain information about their localizations.  Further on, the
inference method for XY pairwise models can be applied to the
optimization of the output signal from complex random media
\cite{Popoff10,Akbulut11}, including disordered optical fibers, by
inferring the elements of the transmission matrix.

Further investigation on inference of waves can include alternative probes of the linear problem here reported by means of  other inference methods such as pseudo-likelyhood.
Most interesting is the generalization to
nonlinear problems, allowing for the reconstruction of the properties of light modes in both ordered and random lasers, both in the
continuous and in the pulsed regime \cite{Marruzzo14,Antenucci14b}.

\section*{Acknowledgments}
We thank Marco Zamparo and Riccardo
Zecchina for interesting discussions on the problem.
The research leading to these results has
received funding from the Italian Ministry of Education, University
and Research under the Basic Research Investigation Fund (FIRB/2008)
program/CINECA grant code RBFR08M3P4 and under the PRIN2010 program,
grant code 2010HXAW77-008 and from the People Programme (Marie Curie
Actions) of the European Union's Seventh Framework Programme
FP7/2007-2013/ under REA grant agreement n¡ 290038, NETADIS project.
\section*{Bibliography}

\providecommand{\newblock}{}

\end{document}